\documentclass{article}
\usepackage{amssymb}
\usepackage{amsmath}
\usepackage{graphicx}
\usepackage{dcolumn}
\usepackage{bm}
\usepackage[round]{natbib}

\newcommand{\pc}{\pb_{c}}	
\newcommand{\pz}{p_{0}}	
\newcommand{\pb}{p}	
\newcommand{\pw}{{\bar{p}}}	
\newcommand{\pest}{p^*}		
\newcommand{\mec}{\chi}		
\newcommand{\mecw}{\bar{\chi}}		
\newcommand{\mecb}{\chi}		
\newcommand{\EC}{\mathcal{X}}		
\newcommand{\ECw}{\mathcal{X}}		
\newcommand{\convex}{\mathcal{K}}		
\newcommand{\convexring}{\mathcal{R}}		
\newcommand{\patternw}{\bar{P}}		
\newcommand{\patternb}{P}		
\newcommand{\lattice}{\Lambda}		
\newcommand{\cell}{C}		

\newcommand{\EQ}[1]{Equation~\ref{eq:#1}}

\newcommand{\FIG}[1]{Figure~\ref{fig:#1}}
\newcommand{\TAB}[1]{Table~\ref{tab:#1}}

\begin{document}
\title{Supplementary Material\\ \emph{Topological estimation of percolation thresholds}}

\author{Richard A. Neher, Klaus Mecke and Herbert Wagner}

\maketitle

\section*{A combinatorial approach to the Euler Characteristic}
In the main body of this paper, we introduced the Euler characteristic (EC) and calculated its mean value
per site for percolation clusters on two dimensional lattices. 
The approach presented in the main text can be generalized to higher dimensional
cell complexes where the EC is given by the alternating sum of the number of cells of different dimension. 

Here, we want to present a complementary definition not restricted to cell complexes and better suited
to study continuum percolation.  Following the 
combinatorial approach \citep{Hadwiger_55}, we start with the family $\convex^{d}$ of closed, bounded,
and convex subsets $A\subset \mathbb{R}^{d}$, $0\leq\dim(A)\leq d$, and define the EC on $\convex^{d}$ by
\begin{equation}
\label{eq:EC_convex}
\ECw(A):=\biggl\lbrace 
\begin{array}{l}
1,\:A\in\convex^{d}\backslash \emptyset\:,\\
0,\:A=\emptyset\in \convex^{d}\:.
\end{array}
\end{equation}
Next, we construct a family $\convexring^{d}$ of spatial patterns $\patternb$ by forming finite unions
of convex bodies $A_i\in \convex^{d}$, i.e. $\patternb=\bigcup_{i=1}^{n}A_i$, and promote the EC to a valuation
on the convex ring $\convexring^{d}$ by requiring additivity \citep{Hadwiger_55}
\begin{equation}
\label{eq:ec_additivity}
\ECw(A\cup B):=\ECw(A)+\ECw(B)-\ECw(A\cap B).
\end{equation}
The value of $\ECw(A\cup B)$ is well-defined, since $A,B\in \convex^{d}$ implies $A\cap B\in\convex^{d}$.
Iteration of \EQ{ec_additivity} leads to 
\begin{equation}
\label{eq:ec_convexring}
\ECw(\bigcup_{i=1}^{n}A_i)=\sum_{k=1}^{n}(-1)^{k-1}\sum_{i_1<i_2\ldots<i_k}\ECw(A_{i_1}\!\cap\! A_{i_2}\ldots \cap
\! A_{i_k}),
\end{equation}
for $A_1, \dots, A_n\in \convex^{d}$.
\citeauthor{Hadwiger_55} has shown that the combinatorial EC agrees with the 
Euler-Poincar\'e characteristic from algebraic topology, when the latter is restricted 
to the convex ring $\convexring^{d}$.
It is now straightforward to turn these formal concepts into practical tools for the study of percolation 
clusters. 

\begin{figure}
\centering
\includegraphics[width=0.45\columnwidth]{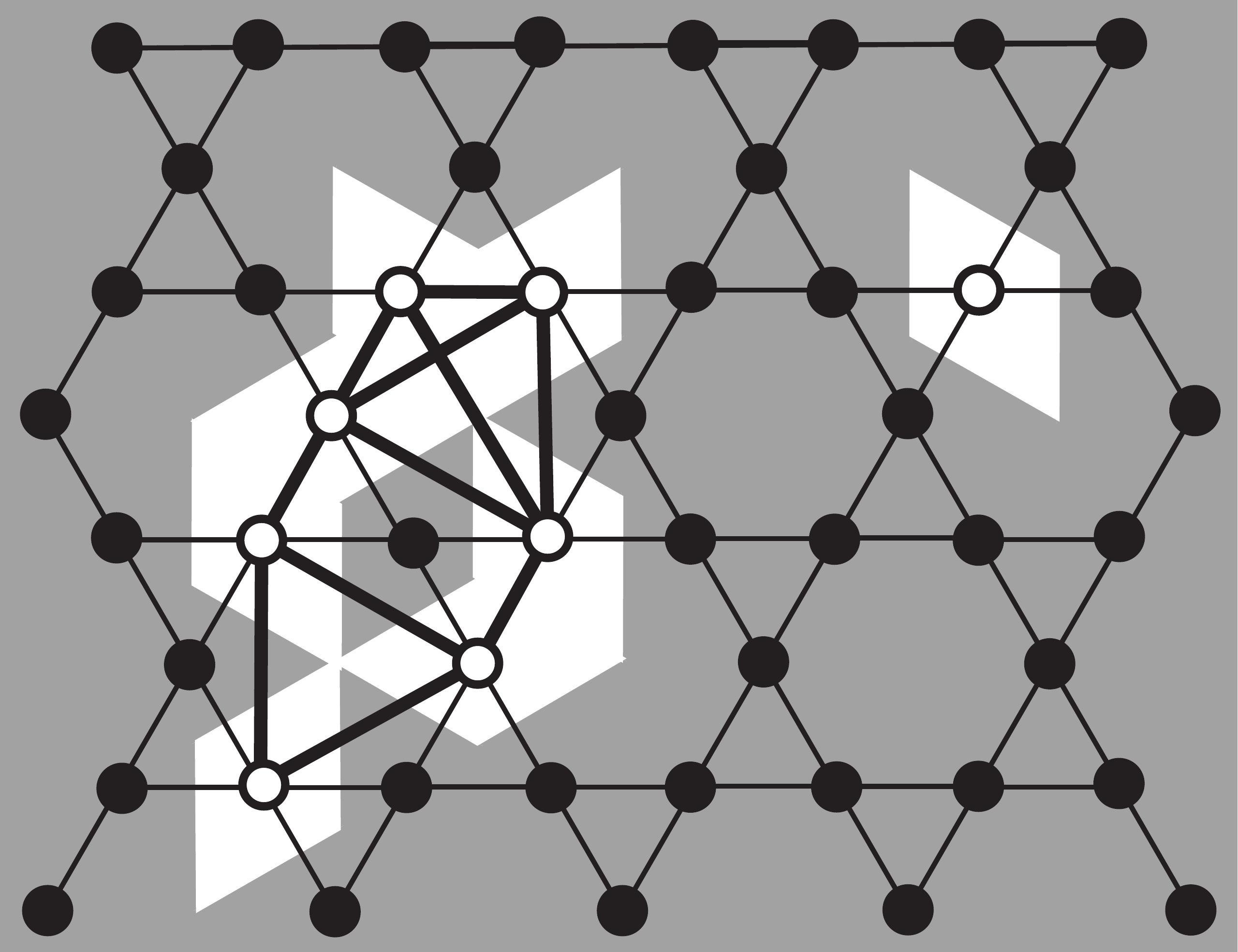}
\hspace{0.2cm}
\includegraphics[width=0.45\columnwidth]{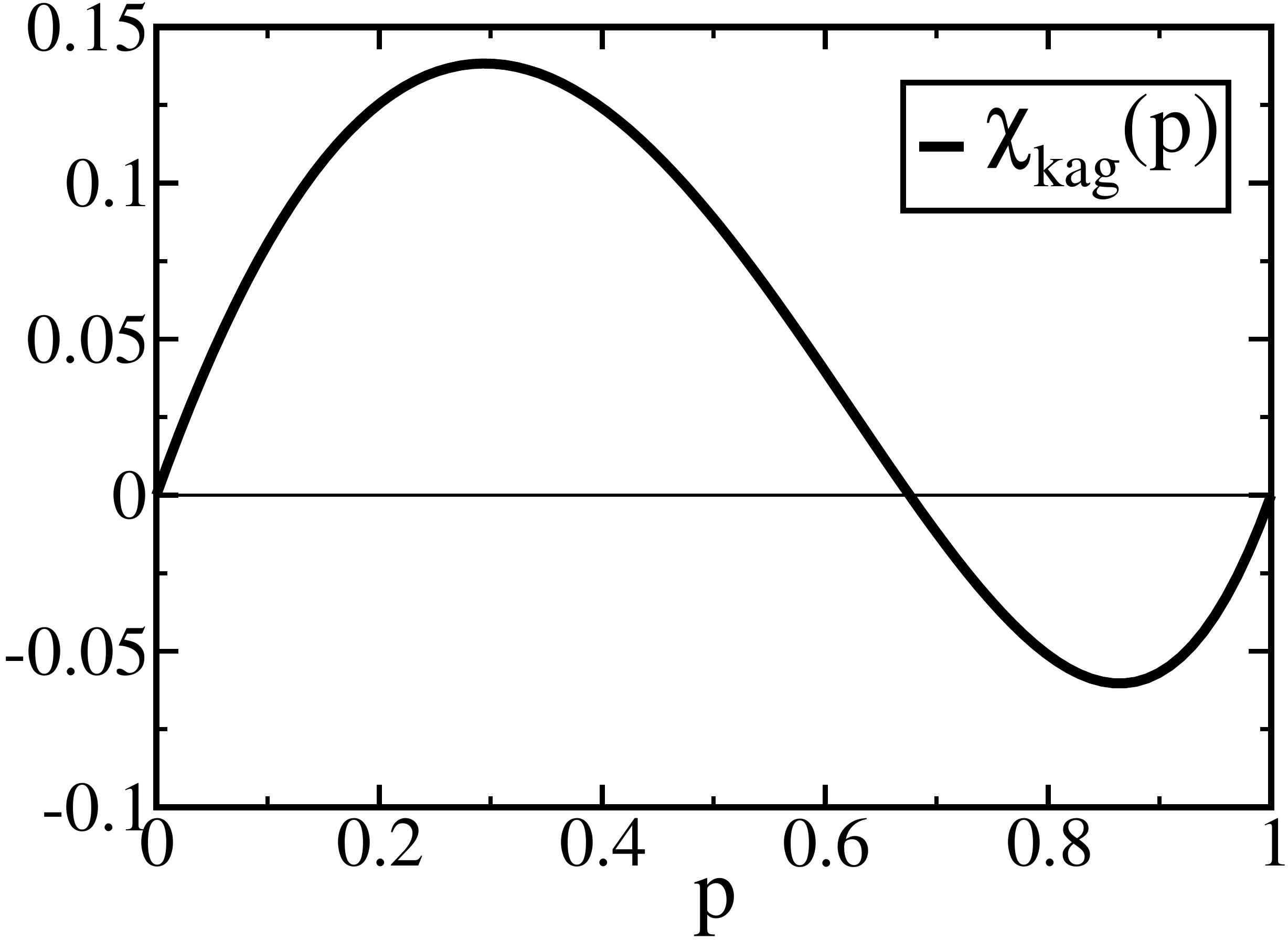}
\caption{\label{fig:supp_lattice_clusters} Clusters on the kagom\'e lattice. 
Left: To each white vertex, we attach a face of the dual lattice.
White faces are connected if they share a dual edge or vertex, as indicated by the thick black lines. 
The white pattern displayed consists of two clusters $\patternw_1$ and $\patternw_2$ with 
$size(\patternb_1)=7$ and $size(\patternb_2)=1$.
The EC of the pattern is given by $\ECw(\patternw_1\cup\patternw_2)=\ECw(\patternw_1)+\ECw(\patternw_2)=0+1$.
Right: The MEC of white clusters for occupation probability $\pw$.}
\end{figure}

At first, we want to reproduce the relations presented in the main text. 
To this end, consider site percolation on a finite planar 2-d lattice $\lattice$ with periodic boundary 
conditions and let each lattice vertex be white with probability $\pw$ and black
with probability $\pb=1-\pw$
\footnote{Within the combinatorial approach, the connectivities of white vertices are more natural than the connectivities of 
black vertices; we therefore start out with a white pattern and recover the black patterns by complementation later.}. 
To construct spatial patterns corresponding to the white vertices, 
we attach to each white vertex a face of the
of the dual lattice $\lattice^{*}$, as illustrated in \FIG{supp_lattice_clusters} for the kagom\'e
lattice. The $k$-cells $\cell_k^{*}$ of $\lattice^{*}$, i.e. its vertices $v^{*}$, edges $e^{*}$ and faces $f^{*}$ 
for $k=0,1,2$, are convex bodies as introduced previously and the EC of union of cells
can be calculated using \EQ{ec_convexring}.
A white cluster is now defined as a maximal collection of white faces $f^{*}$ with nonempty 
pairwise intersections. Since a face is closed, meaning its boundary $\partial f^{*}$ is part
of the face ($\partial f^{*}=f^{*}\cap \partial f^{*}$), two faces are connected if they
share an edge $e^{*}$ or only a vertex $v^{*}$. 
For example, the configuration of white cells
on the kagom\'e lattice shown in \FIG{supp_lattice_clusters} forms a pattern 
of two clusters $\patternw_1$ and $\patternw_2$. 

In percolation theory, the occurrence of a pattern $\patternw=\cup_{i=1}^{n_2^{*}}f_i^{*}\in \convexring^{2}$
is a probabilistic event. Therefore, $\ECw(\patternw)$ is now a random variable. Its mean value per
site of the lattice $\lattice$ is defined by
\begin{equation}
\label{eq:mec_definition}
\mecw(\pw)=\lim_{N\to \infty} \frac{1}{N}\langle \ECw\left(\cup_{i=1}^{n_2^{*}}f_i^{*}\right)\rangle_\pw\;,
\end{equation}
where $N$ is the number of vertices of $\lattice$. To evaluate the average of the EC, we
use the iterated additivity relation from \EQ{ec_convexring} with the $A_i$ replaced by $f_i^{*},\; i=1,\ldots, n_2^{*}$.
The task is simplified by writing a \emph{closed} polyhedral face $f^{*}$ as the union of its
\emph{disjoint} open components (open interior $\check{f}^{*}$ together with the open edges $\check{e}^{*}$
and the vertices $\check{v}^{*}$ from the boundary $\partial f^{*}$), and adopting the definition 
\citep{Nef_81, Likos_95}
\begin{equation}
\label{eq:ec_open_cells}
\ECw(\check{C_k^{*}})=(-1)^{k},\; k=0,1,2,
\end{equation}
for the EC of open $k$-dimensional lattice cells.
With this representation of a white pattern, we obtain
\begin{eqnarray}
\label{eq:open_cells}
\nonumber
\ECw\left(\cup_{i=1}^{n_2^{*}}f_i^{*}\right)&=&\sum_{i=1}^{n_2^{*}}\ECw(\check{f}_i^{*})+\sum_{i=1}^{n_1^{*}}\ECw(\check{e}_i^{*})+\sum_{i=1}^{n_0^{*}}\ECw(\check{v}_i^{*})\\
&=&n_2^{*}-n_1^{*}+n_0^{*}=n_0-n_1+n_2.
\end{eqnarray}
Here $n_k^{*},\; k=0,1,2$ denotes the number of white faces, edges and vertices of the pattern 
on $\lattice^{*}$ and the last equality follows from duality.
The average of the EC depends on the specific geometry of the underlying lattice. 
As an example, let us consider the eleven Archimedian lattices (comp. main text, \citep{Gruenbaum_86}).
In an Archimedian lattice with vertex type $\left(n_1, \ldots, n_z\right)$ and periodic boundary conditions, the fractional numbers of 
bonds and polygonal faces with $n_i$ edges per site of the lattice $\lattice$ are given 
by $\frac{z}{2}$ and $\frac{1}{n_i}$, respectively. With this structural information, we find the averages
\begin{eqnarray}
\label{eq:archi_cells}
\nonumber
\frac{1}{N}\langle n_2^{*}\rangle&=&\pw,\\
\frac{1}{N}\langle n_1^{*}\rangle&=&\frac{z}{2}(1-\pb^{2}),\\ \nonumber
\frac{1}{N}\langle n_0^{*}\rangle&=&\sum_{i=1}^{z}\frac{1}{n_i}(1-\pb^{n_i}).
\end{eqnarray}
These expressions are obtained by noting that a face $f_i^{*}$ is white with probability $\pw=1-\pb$;
an edge $e_i^{*}$ is white with probability $1-\pb^{2}$, since it belongs to a white pattern if at least one 
of the faces sharing $e_i^{*}$ is white; likewise, a vertex $v_i^{*}$ is white, if at least one of the faces 
sharing $v_i^{*}$ is white, and this happens with probability $1-\pb^{n_i}$. Thus from Eq.~\ref{eq:ec_open_cells},
\ref{eq:open_cells} and \ref{eq:archi_cells}, we arrive at the result
\begin{equation}
\label{eq:ec_archi}
\mecw(\pw)=p-\left[1-(1-\pw)^{2}\right]\frac{z}{2}+\sum_i\frac{a_i}{n_i}\left[1-(1-\pw)^{n_i}\right]
\end{equation}
for the MEC of percolation patterns on Archimedian lattices. For example, \EQ{ec_archi} yields for the 
kagom\'e lattice $(3,6,3,6)$
\begin{equation}
\label{eq:ec_kagome}
\mecw_{kag}(\pw)=\pw(1-\pw)\left(1-4\pw+\frac{10}{3}\pw^{2}-\frac{5}{3}\pw^{3}+\frac{1}{2}\pw^{4}\right),
\end{equation}
and for the triangular lattice
\begin{equation}
\label{eq:ec_triangular}
\mecw_{tri}(\pw)=\pw(1-\pw)\left(1-2\pw\right).
\end{equation}
A given white pattern $\patternw=\cup_{j=1}^{n_2^{*}}f_j^{*}$ determines a unique black pattern
$\patternb$, as the set-complement, $\patternw_{comp}$, of $\patternw$ with respect to the lattice $\lattice$, and
we have
\begin{equation}
\label{eq:ec_complement}
\EC(\patternb\cup\patternw)=\EC(\patternb)+\EC(\patternw)=0.
\end{equation}
The first equality holds because $\patternb\cap\patternw=\emptyset$ by definition of the 
complement, the second equality follows since $\patternb\cup\patternw$ is topologically a two dimensional
torus.

Obviously, the complementary pattern $\patternb$ may be viewed as an aggregate
of percolation clusters of faces of $\lattice^{*}$ chosen with probability $\pb=1-\pw$ and colored black.
The MEC of black patterns is defined by
\begin{equation}
\label{eq:mec_white}
\mecb(\pb)=\lim_{N\to\infty} \frac{1}{N}\langle \EC(\patternb)\rangle_{\pb}.
\end{equation}
Consequently, \EQ{ec_complement} implies the symmetry-type relation
\begin{equation}
\label{eq:ec_symmetry}
\mecw(1-p)+\mecb(p)=0.
\end{equation}
Thus, the MEC of black patterns is obtained from that of white patterns via
simple substitution.  After substitution, we obtain exactly the same polynomials for $\mecb(\pb)$ as in the main text, for example:
\begin{eqnarray}
\label{eq:ec_white_examples}
\nonumber
\mecb_{kag}(\pb)&=&\pb(1-\pb)\left(1-\pb-\frac{1}{3}(\pb^{2}+\pb^{3}+\pb^{4})\right),\\
\mecb_{tri}(\pb)&=&\pb(1-\pb)(1-2\pb)=\mecw_{tri}(\pb)
\end{eqnarray}
We note, that the trianglular lattice is the only one among the Archimedian lattice with equal 
``black'' and ``white'' MECs. 

\newpage
\section*{Summary of two-dimensional lattices}
In this section, we collect all formulae for the mean Euler characteristic of the various
lattices discussed in the main text and present tables containing the numerical values of $\pz$, 
$\pest$, and $\pc$. We also discuss some irregular lattices and the dual 
Archimedean lattices, which are not included in the main paper. \textrm{Maple}$^{\textrm{TM}}$ sheets used to calculate the
MECs as well as numerical values of $\pz$ and $\pest$ are available from the authors on request. 

\subsection*{Archimedean lattices -- site percolation}
Each vertex of an Archimedean lattice is surrounded by $z$ regular polygons with $n_1,\ldots, n_z$ edges. In terms of the $n_i$, the MEC of an Archimedian lattice 
is given by 
\begin{equation}
 \mecb(\pb)=\pb(1-\pb)\left( 1-\pb\sum_{i=1}^{z}\frac{1}{n_{i}}\sum_{\mu=0}^{n_{i}-3}\pb^\mu\right).
\end{equation}
The vertex configurations, and a comparison of the zero crossing, our estimator of $\pc$, and the 
actual threshold are given in \TAB{supp_archimed}.
\begin{table}[hb!]
\centering
\begin{tabular}{|l||r|r||r|}
\hline
$n_1,\ldots, n_z$ & $\pz$ &$\pest$&$\pc^{site}$ \\

\hline
    $3,12,12$ &$0.8395$ &$0.7869$&$0.807900\ldots$ \\ \hline
    $4,6,12$&$0.7833$ &$0.7373$ &$0.747806(4)$\\ \hline
    $4,8,8$&$0.7689$ &$0.7269$ &$0.729724(3)$\\ \hline
    $6,6,6$ &$0.7413$&$0.7043$ &$0.697043(3)$\\ \hline
    $3,6,3,6$ &$0.6756$&$0.6462$ &$0.6527036\ldots$ \\ \hline
    $3,4,6,4$ &$0.6468$&$0.6224$ &$0.621819(3)$ \\ \hline
    $4,4,4,4$ &$0.6180$&$0.5987$ &$0.5927460(5)$\\ \hline
    $3,3,3,3,6$  &$0.5913$&$0.5752$  & $0.579498(3)$\\ \hline
    $3,3,4,3,4$ &$0.5616$ &$0.5511$ &$0.550806(3)$ \\ \hline
    $3,3,3,4,4$ &$0.5616$&$0.5511$ &$0.550213(3)$  \\ \hline
    $3,3,3,3,3,3$ &$0.5$&$0.5$&$0.5$\\ \hline
\end{tabular}
\caption{\label{tab:supp_archimed} Archimedean lattices -- site percolation.  Lattices are 
ordered with decreasing $\pc$, the percolation thresholds are from \cite{Suding_PRE_99} and references therein. These values are plotted in Figure 3 of the main paper.}
\end{table}

\subsection*{Archimedean lattices -- bond percolation}
The MEC of the covering lattice of an Archimedean lattice with vertex type $n_1,\ldots, n_z$ is given by
\begin{equation}
\mecb(\pb)=-\pb+\frac{2}{z}\left(1-(1-\pb)^z\right)+\sum_i\frac{2}{zn_i}\pb^{n_i}.
\end{equation}
\begin{table}
\centering
\begin{tabular}{|l||r|r||r|}
\hline
$n_1,\ldots, n_z$& $ \pz$&$\pest$&$p_c^{bond}$ \\ \hline
\hline
$3,12,12$ &  $0.7580$&$0.7098$&$0.7404219(8)$ \\ \hline
$4,6,12$ &  $0.7054$&$0.6685$&$0.6937338(7)$ \\ \hline
$4,8,8$ &$0.6964$&$0.6623$&$0.6768023(6)$ \\ \hline
$6,6,6$ & $0.6756$&$0.6462$&$0.6527036\ldots$ \\ \hline
$3,6,3,6$ &$0.5277$&$0.5227$&$0.5244053(3)$ \\ \hline
$3,4,6,4$ &$0.5134$&$0.5111$&$0.5248325(5)$ \\ \hline
$4,4,4,4$& $0.5$&$0.5$&$0.5$ \\ \hline
$3,3,3,3,6$&$0.4069$&$0.4233$&$0.4343062(5) $ \\ \hline
$3,3,4,3,4$ &$0.3992$&$0.4166$&$0.4141374(5) $ \\ \hline
$3,3,3,4,4$ &$0.3992$&$0.4166$&$0.4196419(4) $ \\ \hline
$3,3,3,3,3,3$&$0.3244$&$0.3538$&$0.3472963\ldots$ \\ \hline
\end{tabular}
\caption{Archimedean lattices -- bond percolation. Lattices are in the same order as in \TAB{supp_archimed}, numerical estimates of percolation thresholds are from \cite{Parviainen_JPA_07}.
This data is plotted in Figure 3 of the main paper.}
\label{tab:supp_archibond}
\end{table}

\subsection*{Dual Archimedean lattices -- site percolation}
\begin{figure}
\centering
\includegraphics[width=0.6\columnwidth]{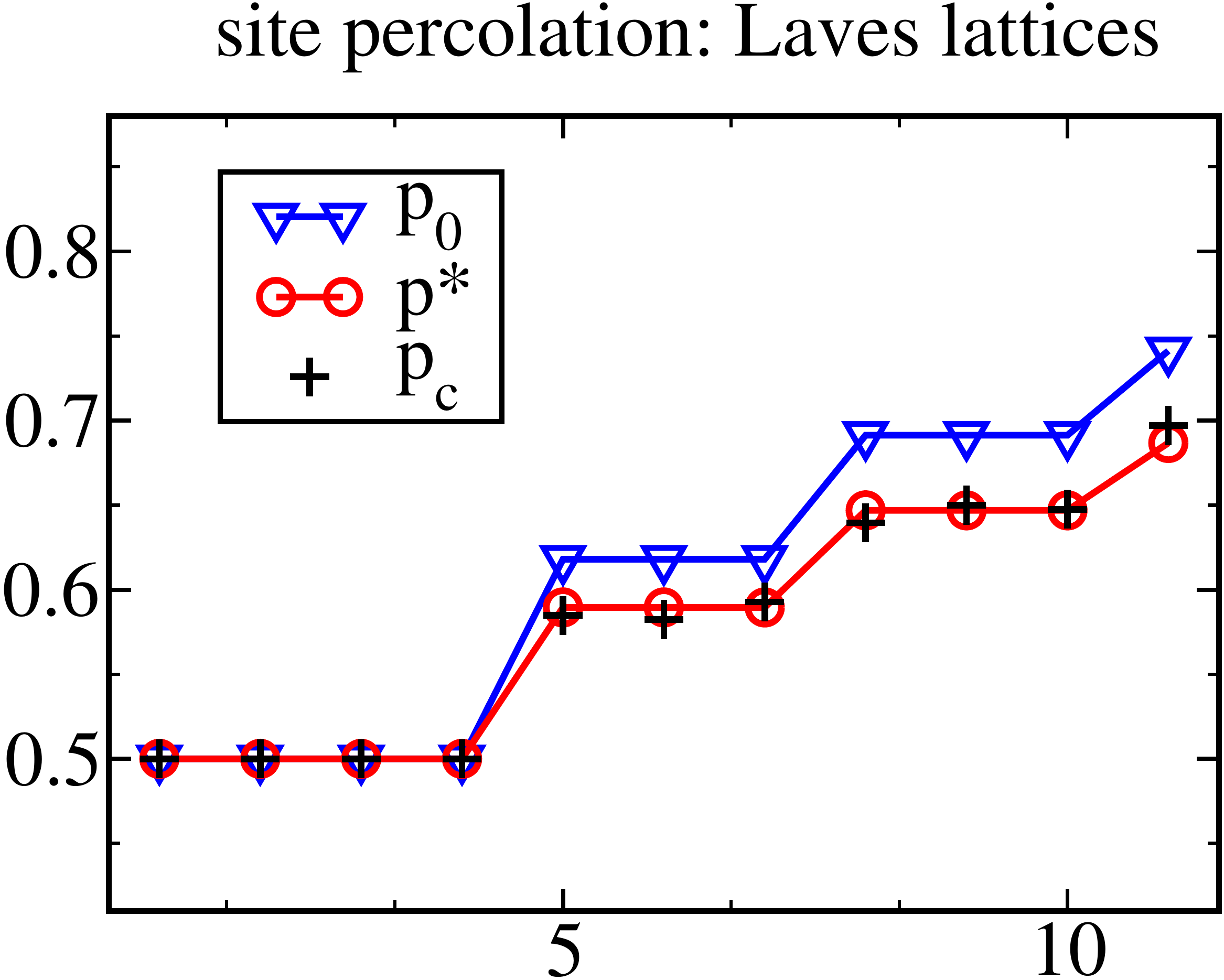}
\caption{Dual Archimedean lattices -- site percolation. 
The solution of Equation 18 of the main text, $\pest$, is a good estimate of $\pc^{site}$ for all dual Archimedian 
lattices, also known as Laves lattices. For numerical values of $\pz$, $\pest$ and $\pc$, see \TAB{supp_archimed_dual}.}
\label{fig:supp_archimed_dual}
\end{figure}

The MEC of the dual lattice of an Archimedean lattice depends only on the
coordination number of the Archimedean lattice. 
\begin{equation}
\label{eq:mec_dual_archimed}
\mecb(\pb)=\pb(1-\pb)\left(1-\frac{2}{z-2}\sum_{i=1}^{z-2}\pb^i\right)
\end{equation}
A graph illustrating the relationship between $\pc$ and the zero crossing of the MEC and 
the performance of our estimator $\pest$ is shown in \FIG{supp_archimed_dual}. 
The dual Archimedean lattices are also known as Laves lattices.

\begin{table}[h]
\centering
\begin{tabular}{|l||r|r||r|}
\hline
dual($n_1,\ldots, n_z$)& $ \pz$&$\pest$&$p_c^{site}$ \\ \hline
\hline
dual($3,12,12$) &  $0.5$&$0.5$&$0.5$ \\ \hline
dual($4,6,12$) &  $0.5$&$0.5$&$0.5$ \\ \hline
dual($4,8,8$) &$0.5$&$0.5$&$0.5$ \\ \hline
dual($6,6,6$) & $0.5$&$0.5$&$0.5$ \\ \hline
dual($3,6,3,6$) &$0.6180$&$0.5987$ & 0.5848(2)\\ \hline
dual($3,4,6,4$) &$0.6180$&$0.5987$ &$0.5824(2)$ \\ \hline
dual($4,4,4,4$) &$0.6180$&$0.5987$ &$0.5927460(5)$ \\ \hline
dual($3,3,3,3,6$)&$0.6914$&$0.6612$&$0.6396(2)$ \\ \hline
dual($3,3,4,3,4$) &$0.6914$&$0.6612$&$0.6500(2)$ \\ \hline
dual($3,3,3,4,4$) &$0.6914$&$0.6612$&$0.6476(2)$ \\ \hline
dual($3,3,3,3,3,3$)&$0.7413$&$0.7043$&$0.697043(3)$ \\ \hline
\end{tabular}
\caption{Dual Archimedean lattices -- site percolation. Lattices are in the same order as in \TAB{supp_archimed}, numerical estimates of percolation thresholds are from \citep{Marck_03}; see
\FIG{supp_archimed_dual} for a plot of this data.}
\label{tab:supp_archimed_dual}
\end{table}

\subsection*{2-uniform lattices -- site percolation}
\label{sec:2-uniform}
While Archimedean lattice have only one vertex type, the vertices of 2-uniform lattices are of two 
distinct types. A vertex is again characterized
by the polygons surrounding it. Each vertex type $i$ makes up a fraction $s_i$ of the total vertices
and a 2-uniform lattice is typically denoted by $s_1(n_1^1,\dots, n_{z_1}^1)+s_2(n_1^2,\dots, n_{z_2}^2)$. The MEC can be expressed in terms of the vertex configuration:
\begin{equation}
  \mecb(\pb)=\pb(1-\pb)\left( 1-\pb\sum_{\nu=1,2}\sum_{i=1}^{z_\nu}\frac{s_\nu}{n_{i}^\nu}\sum_{\mu=0}^{n_{i}^\nu-3}\pb^\mu \right).
\end{equation}
We determined the percolation thresholds of all 2-uniform lattices using the algorithm by
\citeauthor{Newman_PRE_01} \citep{Newman_PRE_01}. Simulation were performed 
on lattices with linear dimension $L=64$, 128, 256, 512 and 1024. The probabilities at which
a spanning appears on a finite lattice were extrapolated to infinite lattice size using the well known 
finite size scaling relations. 
\begin{table}[hb!]
\centering
\begin{tabular}{|l||r|r||r|}
\hline
 $s_1(n_1^1,\dots, n_{z_1}^1)+s_2(n_1^2,\dots, n_{z_2}^2)$& $\pz$ &$\pest$&$\pc^{site}$ \\
\hline
\hline
$\frac{1}{2}(3,4,3,12)+\frac{1}{2}(3,12^2)$&$0.7909$&0.7403&$0.7680(2)$\\
$\frac{1}{3}(3,4,6,4)+\frac{2}{3}(4,6,12)$&$0.7424$&0.7016&$0.7157(2)$\\
$\frac{1}{7}(3^6)+\frac{6}{7}(3^2,4,12)$&$0.6918$&0.6555&$0.6808(2)$\\
$\frac{2}{3}(3^2,6^2)+\frac{1}{3}(3,6,3,6)$&$0.6756$&0.6462&$0.6499(2)$\\
$\frac{1}{7}(3^6)+\frac{6}{7}(3^2,6)$&$0.6532$&0.6270&$0.6329(2)$\\
$\frac{4}{5}(3,4^2,6)+\frac{1}{5}(3,6,3,6)$&$0.6526$&0.6271&$0.6286(2)$\\
$\frac{4}{5}(3,4^2,6)+\frac{1}{5}(3,6,3,6)$&$0.6526$&0.6271&$0.6279(2)$\\
$\frac{2}{3}(3,4^2,6)+\frac{1}{3}(3,4,6,4)$&$0.6468$&0.6224&$0.6221(2)$\\
$\frac{1}{2}(3^4,6)+\frac{1}{2}(3^2,6^2)$&$0.6354$&0.6119&$0.6171(2)$\\
$\frac{1}{2}(3^3,4^2)+\frac{1}{2}(3,4,6,4)$&$0.6053$&0.5874&$0.5885(2)$\\
$\frac{1}{2}(3^2,4,3,4)+\frac{1}{2}(3,4,6,4)$&$0.6053$&0.5874&$0.5883(2)$\\
 $\frac{1}{2}(3^3,4^2)+\frac{1}{2}(4^4)$& $0.5907$&0.5755&$0.5720(2)$\\
$\frac{2}{3}(3^3,4^2)+\frac{1}{3}(4^4)$&$0.5811$&0.5675&$0.5648(2)$\\
$\frac{1}{4}(3^6)+\frac{3}{2}(3^4,6)$ &$0.5684$&0.5563&$0.5607(2)$\\
$\frac{1}{2}(3^3,4^2)+\frac{1}{2}(3^2,4,3,4)$&$0.5616$&0.5511&$0.5505(2)$\\
$\frac{1}{3}(3^3,4^2)+\frac{2}{3}(3^2,4,3,4)$&$0.5616$&0.5511&$0.5504(2)$\\
$\frac{1}{7}(3^6)+\frac{6}{7}(3^2,4,3,4)$&$0.5530$& 0.5439&$0.5440(2)$\\
$\frac{1}{2}(3^6)+\frac{1}{2}(3^4,6)$&$0.5453$&0.5374&$0.5407(2)$\\
$\frac{1}{3}(3^6)+\frac{2}{3}(3^3,4^2)$&$0.5414$&0.5343&$0.5342(2)$\\
$\frac{1}{2}(3^6)+\frac{1}{2}(3^3,4^2)$&$0.5311$&0.5258&$0.5258(2)$\\
\hline
\end{tabular}
\caption{\label{tab:supp_2uniform}2-uniform lattices -- site percolation. Lattices are ordered
according to decreasing percolation threshold. Compare with Figure 3 of the main paper.}
\end{table}

\subsection*{Irregular lattices}
\begin{figure}
\centering
\includegraphics[width=0.6\columnwidth]{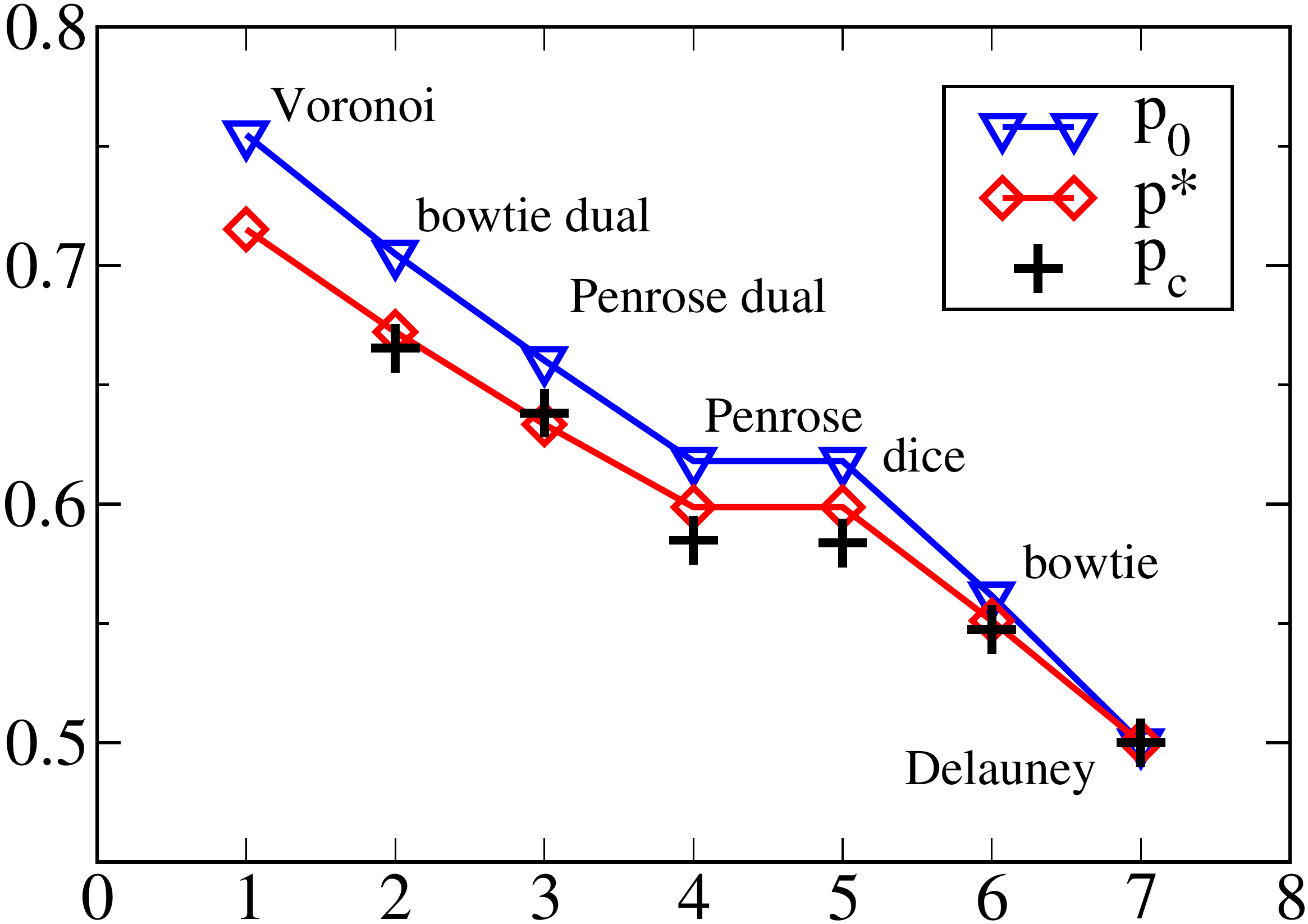}
\caption{Irregular lattices -- site percolation. The relationship between $\pz$, $\pest$ and $\pc^{site}$ also holds for irregular, random and quasi-periodic
lattices. For numerical values, see \TAB{supp_irreg}.}
\label{fig:supp_irreg}
\end{figure}

For lattices that are not included in any of the above classes
classes, we present the polynomials $\mecb(\pb)$ here. \TAB{supp_irreg} and \FIG{supp_irreg}
 summarize our results for irregular lattices.

In addition to the Archimedian lattices, a number of other lattices have been studied by 
\cite{Suding_PRE_99}, including the \emph{bowtie} lattice with vertex configuration
$\frac{1}{2}\left(3^4, 4^2\right)+\frac{1}{2}\left(3^2, 4^2\right)$ the \emph{dual bowtie}
lattice with vertex configuration $\frac{1}{3}\left(4^2, 6^2\right)+\frac{2}{3}\left(4, 6^2\right)$. 
Given these vertex configurations, $\mecb(\pb)$ can be calculated using the same
formula as for 2-uniform lattices. 

\paragraph{Voronoi tesselation and its dual.}
Consider points randomly and independently distributed on the plane and construct
faces around these points using the Wigner-Seitz construction.
The resulting Voronoi tesselation of the plane consists of faces each of which contains 
exactly one of the randomly distributed points. The faces have six sides on average
and the coordination number of a vertex is three with probability one. The dual of the Voronoi
tesselation is a completely triangulated lattice with the randomly distributed points as vertices. 
The MEC of the dual tesselation is therefore the same as that of the triangular lattice.
To calculate the MEC of the Voronoi lattice, we need to know the distribution $\nu_i$ of the 
faces with different number of sides. This is known approximately \citep{Stoyan_96}
and the MEC is found to be
\begin{equation}
\label{eq:supp_mec_voronoi}
\mecb_{vor}(\pb)=\pb-1+\frac{3}{2}(1-\pb^2)-\frac{1}{2}\sum_{i\geq 3}\nu_i(1-p^i).
\end{equation}
We are not aware of any estimation of the site percolation threshold for the Voronoi
tesselation.

\paragraph{Penrose tiling and its dual.}
The rhombic Penrose tiling is a quasi periodic lattice consisting of 4-sided polygons only. The vertices have
an average coordination number of 4, varying between 3 and 7. The distribution of the 
different vertex types is known \citep{Lu_JStatPhys_87}, but since the tessellation consists
of 4-sided polygons only, the MEC is the same as that of the square lattice. The dual Penrose
tiling, however, has faces with 3 to 7 sides. The coordination number is uniformly $z=4$, and from the
distribution of faces we find for the MEC
\begin{equation}
\label{eq:supp_mec_penrosedual}
\mecb(\pb)=-(1-\pb)+2\left(1-\pb^2\right) -\sum_{i\geq 3}\nu_{i}\left(1-p^{i}\right),
\end{equation}
where $\nu_i$ is the probability that a given face has $i$ sides \citep{Lu_JStatPhys_87}.
The percolation thresholds, $\pz$ and $\pest$ for these lattices are given in \TAB{supp_irreg}.

\begin{table}
\centering
\begin{tabular}{|l|l|r|r||r|}
\hline
lattice &$s_1(n_1^1,\dots, n_{z_1}^1)+s_2(n_1^2,\dots, n_{z_2}^2)$&$\pz$&$\pest$&$\pc^{site}$ \\ \hline
\hline
bowtie &$\frac{1}{2}\left(3,3,3,3,4,4\right)+\frac{1}{2}\left(3,3, 4,4\right)$& $0.5616$ &$0.5511$&$0.5474(8)$ \\ \hline
bowtie dual &$\frac{1}{3}\left(4,4, 6,6\right)+\frac{2}{3}\left(4, 6,6\right)$& $0.7048$&$0.6722$&$0.6649(3)$\\ \hline
dice &$\frac{1}{3}\left(4,4, 4,4,4,4\right)+\frac{2}{3}\left(4, 4,4\right)$& $0.6180$&$0.5987$&$0.5848(2)$\\ \hline
Penrose& &  $0.6180$&$0.5987$&$0.5837(2)$ \\ \hline
Penrose dual &&$0.6602$ & $0.6334$&$0.6381(3)$ \\ \hline
Voronoi &&$0.7548$&$0.7151$&-- \\ \hline
Voronoi dual && $0.5$ & $0.5$&$0.5$ \\ \hline
\end{tabular}
\caption{Irregular lattices -- site percolation. Percolation thresholds are taken from \citep{Marck_PRE_97, Yonezawa_88}. This data is plotted in \FIG{supp_irreg}.}
\label{tab:supp_irreg}
\end{table}

\clearpage
\section*{Three dimensional cubic lattices}
The generalization of  \EQ{ec_open_cells} to three dimensions states that the 
MEC of the white pattern is given by the alternating sum of the mean number of 
cells of different dimensions per vertex. For cubic lattices, natural space filling
cells are given by the Wigner-Seitz cells. While white vertices are connected whenever
their cells share a face, an edge or a vertex, black vertices are connected only when their cells share a face. For the simple cubic lattice, the MEC is given by
\begin{equation}
\mecb_{sc}(\pb) = \pb-3\pb^2+3\pb^4-\pb^8.
\end{equation}

The MEC of the face centered cubic (fcc) lattice is given by
\begin{equation}
\mec^{fcc}(p)=p(1-p)(p^{4}+p^{3}+3p^{2}-5p+1).
\end{equation}
A vertex of the fcc-lattice has 12 nearest neighbors, each of which corresponds to one face of the WSC.

A vertex of the bcc-lattice is connected to eight neighbors. The Wigner-Seitz cell, however, has 14 faces, 
six of which make contact to next nearest neighbors on the lattice. The MEC obtained when using 
the connectivities of the Wigner-Seitz cells corresponds to a lattice where the six next nearest neighbors 
are adjacent.
\begin{equation}
\mec^{bcc}_{14-14}(p)=p(1-p)(6p^{2}-6p+1)
\end{equation}
The MEC of the bcc-lattice for black and white vertices are identical, reflecting the
fact that at any 1-cell (edge) or a 0-cell (vertex) of the WSC no cells meet that do not share a face.


\begin{thebibliography}{12}
\providecommand{\natexlab}[1]{#1}
\providecommand{\url}[1]{\texttt{#1}}
\expandafter\ifx\csname urlstyle\endcsname\relax
  \providecommand{\doi}[1]{doi: #1}\else
  \providecommand{\doi}{doi: \begingroup \urlstyle{rm}\Url}\fi

\bibitem[Hadwiger(1955)]{Hadwiger_55}
H.~Hadwiger.
\newblock {Eulers Charakteristik und kombinatorische Geometrie}.
\newblock \emph{J. Reine und Angew. Math.}, 194:\penalty0 101--110, 1955.

\bibitem[Nef(1981)]{Nef_81}
Walter Nef.
\newblock {Zur Einf\"uhrung der Eulerschen Charakteristik}.
\newblock \emph{Monatshefte f\"ur Mathematik}, 92\penalty0 (1):\penalty0
  41--46, March 1981.

\bibitem[Likos et~al.(1995)Likos, Mecke, and Wagner]{Likos_95}
C.N. Likos, K.R. Mecke, and H.~Wagner.
\newblock Statistical morphology of random interfaces in microemulsions.
\newblock \emph{J.Chem.Phys.}, 102:\penalty0 9350--9361, 1995.

\bibitem[Gr\"{u}nbaum and Shephard(1986)]{Gruenbaum_86}
Gr\"{u}nbaum and Shephard.
\newblock \emph{Tilings and Patterns}.
\newblock W. H. Freeman and Company, 1986.

\bibitem[Suding and Ziff(1999)]{Suding_PRE_99}
P.N. Suding and R.M. Ziff.
\newblock Site percolation thresholds for {Archimedean} lattices.
\newblock \emph{Phys. Rev. E}, 60:\penalty0 275--283, 1999.

\bibitem[Parviainen(2007)]{Parviainen_JPA_07}
Robert Parviainen.
\newblock Estimation of bond percolation thresholds on the {Archimedean}
  lattices.
\newblock \emph{J.~Phys.~A}, 40:\penalty0 9253--9258, 2007.

\bibitem[van~der Marck(2003)]{Marck_03}
S.~van~der Marck.
\newblock The site and bond percolation thresholds of the {Archimedean}
  lattices and their duals.
\newblock private communication, 2003.

\bibitem[Newman and Ziff(2001)]{Newman_PRE_01}
M.E.J. Newman and R.M. Ziff.
\newblock Fast {Monte Carlo} algorithm for site or bond percolation.
\newblock \emph{Phys. Rev. E}, 64:\penalty0 016706, 2001.

\bibitem[D.~Stoyan(1996)]{Stoyan_96}
J.~Mecke D.~Stoyan, W.S.~Kendall.
\newblock \emph{Stochastic Geometry and its Applications}.
\newblock WILEY, 1996.

\bibitem[Lu and Birman(1987)]{Lu_JStatPhys_87}
J~P Lu and J~L Birman.
\newblock Percolation and scaling on a quasi lattice.
\newblock \emph{J. Stat. Phys.}, 48:\penalty0 1057, 1987.

\bibitem[van~der Marck(1997)]{Marck_PRE_97}
S.~van~der Marck.
\newblock Percolation thresholds and universal formulas.
\newblock \emph{Phys. Rev. E}, 55\penalty0 (2):\penalty0 1514--1517, 1997.

\bibitem[Yonezawa et~al.(1988)Yonezawa, Sakamoto, Nos\'e, and
  Hori]{Yonezawa_88}
F.~Yonezawa, S.~Sakamoto, S.~Nos\'e, and M.~Hori.
\newblock Percolation in {Penrose} tiling ans its dual.
\newblock \emph{J. of Non-Crystalline Solids}, 106:\penalty0 262--269, 1988.

\end{thebibliography}
\end{document}